\documentclass[twocolumn,prd,superscriptaddress,showpacs,floatfix,%
preprintnumbers,nofootinbib]{revtex4}
\usepackage{epsf}
\usepackage{dcolumn}
\usepackage{bm}
\usepackage{amssymb}
\usepackage{epsfig}
\usepackage{graphicx}
\usepackage{amsfonts}
\usepackage{amssymb}
\usepackage{latexsym}
\usepackage{citesort}

\long\def\ignore#1{}

\begin{document}

\title{Unparticle constraints from SN~1987A}

\author{Steen Hannestad}
\affiliation{Department of Physics and Astronomy, University of
Aarhus, Ny Munkegade, DK-8000 Aarhus C, Denmark}
\affiliation{Max-Planck-Institut f\"ur Physik
(Werner-Heisenberg-Institut),
F\"ohringer Ring 6, D-80805 M\"unchen, Germany}

\author{Georg Raffelt}
\affiliation{Max-Planck-Institut f\"ur Physik
(Werner-Heisenberg-Institut),
F\"ohringer Ring 6, D-80805 M\"unchen, Germany}

\author{Yvonne Y. Y. Wong}
\affiliation{Max-Planck-Institut f\"ur Physik
(Werner-Heisenberg-Institut),
F\"ohringer Ring 6, D-80805 M\"unchen, Germany}

\date{8 August 2007}

\preprint{MPP-2007-108}

\begin{abstract}
The existence of an unparticle sector, weakly coupled to the standard
model, would have a profound impact on supernova (SN)
physics. Emission of energy into the unparticle sector from the core
of SN~1987A would have significantly shortened the observed neutrino
burst. The unparticle interaction with nucleons, neutrinos, electrons
and muons is constrained to be so weak that it is unlikely to provide
any missing-energy signature at colliders. One important exception are
models where scale invariance in the hidden sector is broken by the
Higgs vacuum expectation value. In this case the SN emission is
suppressed by threshold effects.
\end{abstract}
\maketitle

\section{Introduction}

The notion of unparticles was recently introduced by
Georgi~\cite{Georgi:2007ek,Georgi:2007si} as an interesting
possibility for physics beyond the Standard Model.  In this scenario,
there is a hidden sector with a non-trivial infrared fixed point,
$\Lambda$, below which the sector exhibits scale invariance.  At
energies above $\Lambda$, a hidden-sector operator $O_{\rm UV}$ of
dimension $d_{\rm UV}$ couples to standard model operators $O_{\rm
  SM}$ of dimension $n$ via the exchange of heavy particles of mass
$M$,
\begin{equation}
{\cal L}_{\rm UV} =
\frac{O_{\rm UV} O_{\rm SM}}{M^{d_{\rm UV}+n-4}}\,.
\end{equation}
The hidden sector becomes scale invariant at $\Lambda$. The couplings
then become
\begin{equation}
\label{eq:lu}
{\cal L}_U = C_U \frac{\Lambda^{d_{\rm UV}-d}}{M^{d_{\rm UV}+n-4}}
O_{\rm SM} O_U\,,
\end{equation}
where $O_U$ is the unparticle operator of dimension $d$ in the
low-energy limit, and $C_U$ is a dimensionless coupling constant.
Because of scale invariance, the phase space for $O_U$ resembles that
of $d$ massless particles; the salient feature of the unparticle
sector is that $d$ may take on non-integer values.  Unparticle
phenomenology has been investigated in a large number of recent
papers~\cite{Cheung:2007ue, Luo:2007bq, Chen:2007vv, Ding:2007bm,
  Liao:2007bx, Aliev:2007qw, Li:2007by, Duraisamy:2007aw, Lu:2007mx,
  Stephanov:2007ry, Fox:2007sy, Greiner:2007hr, Davoudiasl:2007jr,
  Choudhury:2007js, Chen:2007qr, Aliev:2007gr, Mathews:2007hr,
  Zhou:2007zq, Ding:2007zw, Chen:2007je, Liao:2007ic, Bander:2007nd,
  Rizzo:2007xr, Cheung:2007ap, Chen:2007zy, Zwicky:2007vv,
  Kikuchi:2007qd, Mohanta:2007ad, Huang:2007ax, Krasnikov:2007fs,
  Lenz:2007nj, vanderBij:2007um, Choudhury:2007cq, Zhang:2007ih,
  Li:2007kj, Nakayama:2007qu, Deshpande:2007jy, Ryttov:2007sr,
  Mohanta:2007uu, Delgado:2007dx, Cacciapaglia:2007jq, Neubert:2007kh,
  Luo:2007me}.

If the unparticle sector indeed appears at low energies in the form of
new massless fields coupled very weakly to standard model particles,
one expects that the usual stellar energy-loss
limits~\cite{Raffelt:1990yz, Raffelt:1999tx}, notably from the
neutrino burst duration of SN~1987A, will provide much more
restrictive limits than any laboratory experiment could achieve.
Invisible axions~\cite{Ellis:1987pk, Raffelt:1987yt, Turner:1987by,
  Hanhart:2000ae, Raffelt:2006cw} and Kaluza-Klein
gravitons~\cite{Cullen:1999hc, Hanhart:2000er, Hannestad:2003yd} are
typical examples where astrophysical limits preclude any realistic
hope of finding these particles directly in collider experiments, and
the same is expected for unparticles as noted by
Davoudiasl~\cite{Davoudiasl:2007jr}. We will here elaborate this idea
in some detail.

If a scalar unparticle operator of dimension $d<2$ couples to the
standard-model Higgs, then scale invariance in the hidden sector is
broken once the Higgs acquires a non-zero VEV so that there is no
unparticle signature at low energies~\cite{Fox:2007sy}.  In this
case the SN~1987A energy-loss argument is moot. Therefore, the
astrophysical limit serves to constrain the possible realizations of
the unparticle concept that could show up as missing energy at
colliders.

The dominant lowest-order process for the emission of new particles
$X$ in a SN core tends to be nucleon bremsstrahlung $N+N\to N+N+X$.
The nucleon density is huge. Their interaction rate is large, and
this process involves the small coupling of the new particles to
lowest order. The emission of unparticles will be fully analogous,
apart from phase-space modifications. In addition, processes of the
form $\bar{f}+f \to U$ now have a nonvanishing phase space and
provide additional contributions. Therefore, the SN~1987A
energy-loss argument provides very restrictive constraints on the
unparticle couplings to nucleons, neutrinos, electrons and muons. We
note that the SN~1987A emission bound is much stronger than other
astrophysical constraints, as is also the case for Kaluza-Klein
graviton emission.

Numerical studies of SN energy losses by axions or Kaluza-Klein
gravitons reveal that the neutrino burst would have been excessively
shortened unless the volume energy-loss rate of the SN core
obeys~\cite{Raffelt:1990yz, Raffelt:1999tx}
\begin{equation}\label{eq:SNcriterion}
Q\alt3\times10^{33}~{\rm erg}~{\rm cm^{-3}}~{\rm s}^{-1}\,,
\end{equation}
where $Q$ is to be calculated at the benchmark conditions $T=30~{\rm
  MeV}$ and $\rho=3\times10^{14}~{\rm g}~{\rm cm}^{-3}$. Armed with
this simple criterion, all that is needed is an estimate of $Q$ for
unparticle emission.

We begin in Sec.~\ref{sec:bremsstrahlung} with the nucleon
bremsstrahlung process, stressing its general features for the
emission of different types of new radiation. In Sec.~\ref{sec:pairs}
we consider the pair-annihilation process for various particles that
are abundant in a SN core. We conclude in Sec.~\ref{sec:discussion}.

\section{Nucleon Bremsstrahlung}
\label{sec:bremsstrahlung}

\subsection{General approach}

A reliable calculation of the bremsstrahlung rate $N+N\to N+N+X$ in a
SN core is not possible because the $NN$ interaction
potential is not well known, $NN$ correlation effects are
likely strong, and multiple-scattering effects can be
important. Therefore, one cannot do much better than an ``educated
dimensional analysis'' based on simple principles.

In the limit $\omega\to0$, where $\omega$ is the frequency of the
emitted radiation, bremsstrahlung factorizes into an essentially
classical radiation process and the ordinary collision.  The advantage
of this approach is that for nucleons one can estimate the collision
rate from measured cross sections or phase shifts even without a
detailed model of the interaction potential~\cite{Hanhart:2000ae,
  Hanhart:2000er}.

Assuming weakly interacting radiation we can employ lowest-order
perturbation theory for the interaction between the radiation and
the medium. The emitted power is then a phase-space integral over
the dynamical structure function $S(\omega,{\bf k})$ of the medium.
Here, $S$ is the thermal average of a correlator of the operator to which
the radiation couples. While we do not know the dynamical structure
function for a strongly interacting medium, we can still take
advantage of its general properties. In particular, the principle of
detailed balancing tells us that the probability for emitting a
quantum of energy $\omega$ from a thermal medium is $e^{-\omega/T}$
times the probability of absorbing one.

These ideas suggest that the volume energy-loss rate is roughly
given by
\begin{equation}
Q\approx\langle\sigma\,v\rangle\,n_B^2\,\int_0^{\infty}d\omega\,
\frac{dI}{d\omega}\Big|_{\hbox{\footnotesize low-$\omega$}}
\,e^{-\omega/T}\,,
\end{equation}
where $\sigma$ is the nucleon scattering cross section, $v$ their
relative velocity, and $n_B$ their number density (the baryon
density). The dimensionless quantity $dI/d\omega$ is the spectrum of
energy emitted in a single collision and includes all coupling
constants and the radiation phase space.

Typically $dI/d\omega$ is of the power-law form $G^2\omega^p$ where
$G$ is a coupling constant of dimension $({\rm energy})^{-p/2}$ and
includes numerical factors. Therefore, we estimate
\begin{equation}\label{eq:Qpower}
Q\approx\langle\sigma\,v\rangle\,n_B^2G^2\,T^{p+1}\,,
\end{equation}
where we have ignored a numerical factor $\Gamma(p+1)$ that plays
little role as long as $p$ is not too large.

\subsection{Graviton Bremsstrahlung}

The usual axion constraints from SN~1987A were derived along these
lines~\cite{Raffelt:2006cw}. Moreover, two of us have carried out this
exercise in detail, including all numerical factors, for the case of
graviton bremsstrahlung~\cite{Hannestad:2003yd}. In this case the
classical emission rate is
\begin{equation}
 \frac{dI}{d\omega}\Big|_{\hbox{\footnotesize low-$\omega$}}
 =\frac{8}{5\pi}\,(G_{\rm N}m^2)\,
 v^4\,\sin^2\Theta_{\rm CM},
\end{equation}
where $G_{\rm N}$ is Newton's constant, $m$ the nucleon mass, $v$ the
velocity of the colliding nucleons, and $\Theta_{\rm CM}$ the
scattering angle in the center-of-mass frame. The quantity $G_{\rm
  N}m^2$ plays the role of the square of a dimensionless coupling
constant and the $v^4$ behavior reflects the quadrupole nature of the
radiation.  Assuming that $NN$ scattering is isotropic with a total
cross section $\sigma$, after performing the nucleon phase-space
integral as well as the $d\omega$ integration it was found that
\begin{equation}\label{eq:Qgrav}
 Q=\left[\frac{512\ln2}{3^{5/2}\,5\,\pi^{3/2}}\,(G_{\rm N}m^2)\,T\right]
 \times\left[\sigma\,n_B^2\,
 \left(\frac{3T}{m}\right)^{5/2}\right]\,.
\end{equation}
Of course it is somewhat arbitrary how to group the numerical
factors. We note that the average velocity of thermal nucleons is
given by $\langle \frac{1}{2}m v^2\rangle=\frac{3}{2}T$ so that
$\langle v^2\rangle=3T/m$, motivating the use of $3T$ rather than $T$
in the last term that derives from the nucleon velocity factors. (Four
powers of velocity come from the quadrupole nature of the radiation,
and another factor from the $NN$ relative velocity.)  In this way the
overall numerical coefficient is 0.818 and thus very close to unity.
Graviton bremsstrahlung has a flat spectrum so that we have $p=0$ in
the spirit of Eq.~(\ref{eq:Qpower}), implying a simple factor
$T^{p+1}=T$ within the first term.

\subsection{Unparticle emission}

In the simplest case unparticles couple via a vector current to
nucleons, an assumption made in all previous studies.
The structure of the coupling is
\begin{equation}
{\cal L}_{UN} = C_{UN}\frac{\Lambda^{d_{\rm UV}-d}}{M^{d_{\rm UV}-1}}
\,\bar{N} \gamma_\mu N\, O_U^\mu\,.
\end{equation}
Nucleon bremsstrahlung would then seem to be normal dipole emission
similar to electromagnetic bremsstrahlung. However, if the interaction
with all nucleons is the same, bremsstrahlung is suppressed in the
nonrelativistic limit, just as it is suppressed in the nonrelativistic
limit for $e^-+e^-\to e^-+e^-+\gamma$.  Therefore, we expect
quadrupole radiation to dominate, involving a factor $v^4$ as in
graviton bremsstrahlung.

Therefore, we scale the emission rate from Eq.~(\ref{eq:Qgrav}),
keeping the second factor that encodes the quadrupole nature and
nucleon phase space, while adapting the first factor to the unparticle
case. From dimensional analysis we thus expect
\begin{equation}\label{eq:QU}
 Q=C_{UN}^2\frac{\Lambda^{2(d_{\rm UV}-d)}}{M^{2(d_{\rm UV}-1)}}
 \,T^{2d-1}\,\sigma\,n_B^2\,
 \left(\frac{3T}{m}\right)^{5/2}\,.
\end{equation}

To evaluate this rate we assume $\sigma=25\times10^{-27}~{\rm
  cm}^2$ that is typical for the relevant conditions and was
recommended in the context of Kaluza-Klein graviton
emission~\cite{Hanhart:2000er}. Applying Eq.~(\ref{eq:SNcriterion}) we
find the constraint
\begin{equation}
C_{UN}\frac{\Lambda^{d_{\rm UV}-d}}{M^{d_{\rm UV}-1}}\,
(30~{\rm MeV})^{d-1}
\alt 3\times10^{-11}\,.
\end{equation}
Assuming $d_{\rm UV}=3$ and $M=1000~{\rm TeV}$, we find
\begin{equation}\label{eq:bounds}
\Lambda \alt \cases{5 \, {\rm GeV} & $d=1$\cr
30 \, {\rm GeV} & $d=3/2$ \cr
900 \, {\rm GeV} & $d=2$},
\end{equation}
where we have set $C_{UN}=1$ for simplicity. These bounds are almost
identical to those derived in Ref.~\cite{Davoudiasl:2007jr} from
scaling the usual axion limits to the unparticle case.  This is not
strictly correct since the coupling structure of axions is different
(axial instead of vector).  Note that if the nucleon-unparticle
coupling was axial the bound should become somewhat stronger because
there would be no $v^4$ quadrupole suppression.

\begin{figure}[t]
\includegraphics[width=7.5cm]{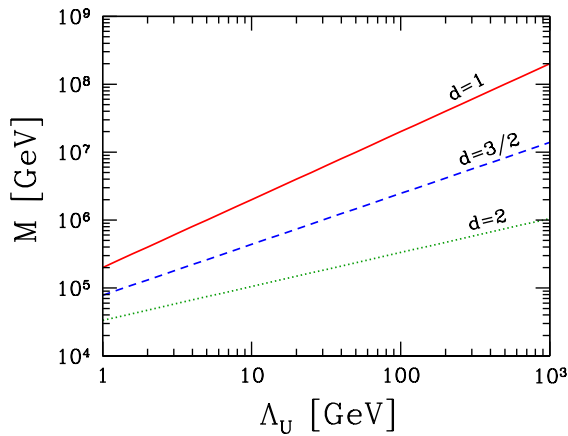}
\caption{Constraints on vector unparticle operators from SN
  bremsstrahlung emission, assuming $d_{\rm UV}=3$, for $d=1$, 3/2,
  and 2 as indicated. The regions below the contours are excluded.
\label{fig:m-lambda}}
\end{figure}

Figure~\ref{fig:m-lambda} shows the same constraints on the
$(\Lambda,M)$-plane.  These should be compared with analogous bounds
from collider experiments in, e.g., Fig.~6 of
Ref.~\cite{Bander:2007nd}.  The SN limits are always far more
restrictive.  As an illustration, for $\Lambda < 1000~{\rm GeV}$,
Ref.~\cite{Bander:2007nd} quotes the constraints $M>7500~{\rm GeV}$
($d=1$), $M>2500~{\rm GeV}$ ($d=3/2$), and $M>1000~{\rm GeV}$ ($d=2$),
to be compared with our much more severe SN limits $M>10^8~{\rm GeV}$
($d=1$), $M>10^7~{\rm GeV}$ ($d=3/2$), and $M>10^6~{\rm GeV}$ ($d=2$).

\section{Pair annihilation}
\label{sec:pairs}

For massless particles $X$, the process $\bar{f}f \to X$ is forbidden
because of energy-momentum conservation.  At next order the process
$\bar{f}+f \to X+X$ is highly suppressed because of the additional
power of the coupling constant. For unparticles, however, the process
$\bar{f}+f \to U$ is allowed and can be a dominant energy-loss process
in a SN core.

The emissivity from pair annihilation of neutrinos is roughly
estimated by
\begin{equation}
Q \sim C_{U\nu}^2 \frac{\Lambda^{2(d_{\rm UV}-d)}}{M^{2(d_{\rm
UV}-1)}} T^{2d+3}\,, \label{eq:pair}
\end{equation}
which leads to the constraints, for $C_{U\nu}=1$,
\begin{equation}
\Lambda \alt \cases{- & $d=1$\cr 20~{\rm GeV} & $d=3/2$ \cr 500~{\rm
GeV} & $d=2$} \label{eq:ann}
\end{equation}
assuming $d_{\rm UV}=3$ and $M=1000~{\rm TeV}$ as before.

For $d=3/2$ and $2$ these bounds are almost identical to those
deduced from bremsstrahlung in Eq.~(\ref{eq:bounds}), although they
are based on different physics.  The reason is that the smaller
neutrino density is compensated by the fact that the $\nu \bar{\nu}$
channel is not suppressed by the $v^4$ factor.  For different values
of $M$ the bounds on $\Lambda$ scale in the same way as those from
the bremsstrahlung calculation so that the results of
Fig.~\ref{fig:m-lambda} apply also in this case.  No bound exists
for $d=1$ from $\bar{f}+f \to U$, since this case corresponds to
annihilation into a single massless particle, and its rate must
vanish exactly because of energy-momentum conservation,

The bounds Eq.~(\ref{eq:ann}) were calculated assuming
non-degenerate, relativistic fermions.  This approximation is very
good for $\nu_\mu$ and $\nu_\tau$.  However, electron neutrinos are
degenerate with a chemical potential $\mu$ typically of order
$150$--$200$~MeV, implying a degeneracy parameter around 5--7 for
$T=30~{\rm MeV}$. The presence of a chemical potential reduces the
number of pairs, i.e., the quantity $n_{\nu}n_{\bar\nu}$ is largest
for a vanishing chemical potential. However, for degeneracy factors
up to roughly 8 the suppression of the annihilation rate is less
than an order of magnitude (for illustration see Fig.~2 of
Ref.~\cite{Buras:2002wt}). The degeneracy factor for electrons is
only slightly larger so that the suppression of $e^+e^-$
annihilation is only slightly worse.

Because of its high temperature the hot proto-neutron star is also
abundant in muons. The muon rest mass of 106 MeV does not lead to a
very substantial difference in number density relative to massless
fermions. At $T=30~{\rm MeV}$ the number density of muons is
suppressed only by a factor $\sim 2.5$ relative to massless fermions,
always ignoring chemical potentials that are small for muons in a SN
core. An additional suppression factor of about 0.5 comes from the
relative velocity of muons so that the annihilation rate overall is
roughly 0.2 times that of a massless fermion species.

Lastly, we note that a bound on the $d=1$ scenario can still be
obtained from pair annihilation of charged leptons based on the
process $e^+ +e^- \to U+ \gamma$ or $\mu^+ +\mu^- \to U+ \gamma$
which is not phase-space suppressed.  Since the squared matrix
element is suppressed only by ${\cal O} (\alpha)$ relative to
$\bar{f}f \to U$, the bound for $d=1$ can be estimated from taking
the emissivity as $\alpha$ times the naive rate given in
Eq.~(\ref{eq:pair}), and taking into account the suppression from
degeneracy (electrons) or the mass threshold (muons). Note that for
$d=1$ the suppression factors enter as a fourth root, giving an
estimated bound for $d=1$ of $\Lambda \alt 20-30$~GeV for
$M=1000~{\rm TeV}$.

Given the overall numerical uncertainties of our limits, we conclude
that the pair annihilation and bremsstrahlung limits are nearly
identical for all cases, i.e., apart from possible difference in the
overall coefficients $C_U$ to different particle species, the
interaction with nucleons, neutrinos of all flavors, electrons and
muons are constrained by the limits shown in Fig.~\ref{fig:m-lambda}.

\section{Discussion}
\label{sec:discussion}

We have applied the well-known energy-loss limit based on the SN~1987A
neutrino burst duration to Georgi's new idea of unparticles that can
manifest themselves as missing energy in collider experiments with a
peculiar phase-space behavior. As expected, the SN limits are very
restrictive as long as the unparticle radiation can be emitted without
threshold at the relatively low energies prevalent in the SN context.
Our approximate constraints shown in Fig.~\ref{fig:m-lambda} apply
without significant modifications to nucleons, neutrinos of all
flavors, electrons and muons.

Unparticle signatures can still be detected at colliders in models
where scale invariance in the hidden sector is broken by the Higgs
vacuum expectation value. In this case the SN emission is
suppressed by threshold effects.
Thus our astrophysical limits provide a severe restriction
on the type of unparticle models that can be detected at colliders.


\begin{acknowledgments}
We thank Alessandro Mirizzi for discussions. This work was supported
in part by the Deutsche For\-schungs\-ge\-mein\-schaft under the
grant No.~TR-27 ``Neutrinos and Beyond'', the Cluster of Excellence
``Origin and Structure of the Universe'' (Munich and Garching), and
by the European Union under the ILIAS project, contract
No.~RII3-CT-2004-506222. SH acknowledges support by the Alexander
von Humboldt Foundation.
\end{acknowledgments}


\raggedright

\end{document}